\documentstyle[12pt,twoside,psfig,epsfig]{article}
\begin{document}
\begin{center}
{\large\bf Collapse of non-spherically symmetric scalar field distributions}
\\[15mm]
Koyel Ganguly$^{*}$ {\footnote{E-mail: koyel\_g\_m@yahoo.co.in}} 
and Narayan Banerjee$^{\dagger}${\footnote {E-mail: narayan@iiserkol.ac.in}}\\
$^{*}$Relativity and Cosmology Research Centre, Department of Physics, 
        Jadavpur University, Kolkata - 700032, India.\\
$^{\dagger}$ IISER - Kolkata, Mohanpur Campus, P.O. BCKV Main Office, District Nadia,\\ West Bengal 741252, India.\\
            
\end{center}
\date{}
\vspace{0.5cm}
{\em PACS Nos.: 04.20 Dw; 04.70 Bw }
\vspace{0.5cm}

\pagestyle{myheadings}
\newcommand{\be}{\begin{equation}}
\newcommand{\ee}{\end{equation}}
\newcommand{\bea}{\begin{eqnarray}}
\newcommand{\eea}{\end{eqnarray}}

\begin{abstract}
In the present work the collapse scenario of some exact non-spherical models with a minimally coupled scalar field is studied. Scalar field collapse with planar as well as toroidal, cylindrical and pseudoplanar symmetries have been investigated. It is shown that the scalar field may have collapsing modes even if it has the equation of state corresponding to that of a dark energy.
\end{abstract}

\section{Introduction}
\par It is now well known that a collapsing matter distribution may not always lead to the formation of a black hole. The spacetime singularity, formed out of the unhindered collapse, may not be covered by an event horizon and thus exposed to an external observer. The formation of black holes, for a spherically symmetric dust distribution, normally depends on the initial density profile\cite{pjoshi}. A star is expected to collapse under its own gravity when its nuclear fuel gets exhausted so that there is no pressure to balance gravity. So the approximation of the matter distribution as dust is quite a well-motivated one. But as there is not much phenomenological example of the equation of state of the high density collapsing fluid, investigations regarding the collapse of other forms of matter become important. Scalar fields naturally step in as a candidate for other forms of matter, mainly because of their ability to mimic different kinds of matter depending on the choice of potentials. A lot of spherical scalar field models already figure in the literature. Some of them involve exact analytical study\cite{christodolou} and some are numerical endeavours\cite{choptuik}. Careful investigations regarding a scalar field collapse bring out the fascinating possibility of a critical phenomenon\cite{gundlach}.
\par Although mathematically elegant, a spherically symmetric collapse, is hardly found in nature as most of the stellar bodies do show departure from sphericity. One may approach the problem in two different ways. One is to look at exact non-spherical models and the other resorts to a perturbative approach where a departure from the spherical symmetry in collapse scenario is considered. For an extensive review, we refer to the work by Miller and Sciama\cite{miller}.
\par The present work investigates some exact non-spherical models with a minimally coupled scalar field. Bronnikov and Kovalchuk made an extensive study of the various non-spherical collapse with a pressureless dust distribution\cite{bronnikov} and then generalized the same to include a charged dust in planar symmetry\cite{brokov} and later an electromagnetic field for various non-spherical distributions\cite{kovalchuk}. The present paper extends the work for scalar fields in planar, cylindrical, toroidal and pseudoplanar symmetries.
\par As the interest in scalar fields has increased manyfold because of their role as a possible candidate as a `dark energy' driving the present accelerated expansion of the Universe(see ref\cite{de} for reviews), the study of the collapse of a scalar field distribution becomes important as this may lead to some idea about the clustering properties of the dark energy which are yet to be ascertained.
\par In section 2 we write down the Einstein's equations for the scalar field distribution for a general kind of metric which gives rise to different symmetries as special cases. Section 3 discusses the collapsing models for a planar symmetry, section 4 describes scalar field collapse for cylindrical, toroidal and pseudoplanar symmetry. In section 5 we make some concluding remarks.
\section{Field Equations}
We take a metric,
$$ds^2=e^{2\gamma}dT^2-e^{2\lambda}d{R}^2-a^2[e^{-2\omega}d{\eta}^2+{f}^2(\eta)e^{2\omega}d{\xi}^2]$$\\
where $\gamma$, $\lambda$, $a$ and $\omega$ depend on $R$ and $T$. For different choices of $f=f(\eta)$ and $\omega$, one 
has different symmetries as,
\begin{center}
$f$ = sin$\eta,$   $\omega = 0$  (spherical), \\
$f$ = sinh$\eta,$   $\omega = 0$  (pseudospherical),\\
$f = 1,$   $ \omega\ne 0$  (pseudoplanar, cylindrical, toroidal),\\
$f = 1,$   $  \omega = 0$  (planar).
\end{center}
In the present work the collapse scenario for the latter two are investigated. The energy momentum tensor `$T_{\mu\nu}$' for scalar field $\phi$ where $\phi$
is a function of `$R$' and `$T$'and $V(\phi)$ is the scalar potential is given as,
\be
T_{\mu\nu}=\phi_{,\mu}\phi_{,\nu}-\left[\frac{1}{2}g^{\alpha\beta}\phi_{,\alpha}\phi_{,\beta}-V(\phi)\right]g_{\mu\nu}
\ee
\par The Einstein equations for the said metric with $T_{\mu\nu}$ given as in (1) with $f=1$ are
\be
2\dot{\gamma}\frac{\dot{a}}{a}e^{-2\gamma}-e^{-2\gamma}(2\frac{\ddot{a}}{a}+\frac{\dot{a}^2}{a^2}+{\dot{\omega}}^{2})
+2{\gamma}^{\prime}\frac{{a}^{\prime}}{a}
e^{-2\lambda} + (\frac{{{a}^{\prime}}^{2}}{a^2}-{{\omega}^{\prime}}^2)e^{-2\lambda}
 = \frac{1}{2}(e^{-2\lambda}{{\phi}^{\prime}}^2 + e^{-2\gamma}{\dot{\phi}}^2)
- V(\phi),
\ee
$e^{-2\gamma}(-\ddot{\lambda}-{\dot{\lambda}}^2-\frac{\ddot{a}}{a}-2\frac{\dot{a}}{a}\dot{\omega}-{\dot{\omega}}^{2}-\ddot{\omega}+\dot{\gamma}\dot{\lambda}
+\dot{\gamma}\frac{\dot{a}}{a}+\dot{\gamma}\dot{\omega}-\dot{\lambda}\frac{\dot{a}}{a}-\dot{\lambda}\dot{\omega})+$
\be
e^{-2\lambda}(\frac{{a}^{\prime\prime}}{a}
+2\frac{{a}^{\prime}}{a}{\omega}^{\prime}+{{\omega}^{\prime}}^2+{\omega}^{\prime\prime}+{\gamma}^{\prime\prime}+{{\gamma}^{\prime}}^2-{\lambda}^{\prime}\frac{{a}^{\prime}}{a}-{\lambda}^{\prime}
{\omega}^{\prime}-{\gamma}^{\prime}{\lambda}^{\prime}+{\gamma}^{\prime}\frac{{a}^{\prime}}{a}+{\gamma}^{\prime}{\omega}^{\prime}) = \frac{1}{2}(-e^{-2\lambda}{{\phi}^{\prime}}^2 + 
e^{-2\gamma}{\dot{\phi}}^2)- V(\phi),
\ee
$e^{-2\gamma}(-\ddot{\lambda}-{\dot{\lambda}}^2-\frac{\ddot{a}}{a}+2\frac{\dot{a}}{a}\dot{\omega}-{\dot{\omega}}^{2}+\ddot{\omega}+\dot{\gamma}\dot{\lambda}
+\dot{\gamma}\frac{\dot{a}}{a}-\dot{\gamma}\dot{\omega}-\dot{\lambda}\frac{\dot{a}}{a}+\dot{\lambda}\dot{\omega})+$
\be
e^{-2\lambda}(\frac{{a}^{\prime\prime}}{a}
-2\frac{{a}^{\prime}}{a}{\omega}^{\prime}+{{\omega}^{\prime}}^2-{\omega}^{\prime\prime}+{\gamma}^{\prime\prime}+{{\gamma}^{\prime}}^2-{\lambda}^{\prime}\frac{{a}^{\prime}}{a}+{\lambda}^{\prime}
{\omega}^{\prime}-{\gamma}^{\prime}{\lambda}^{\prime}+{\gamma}^{\prime}\frac{{a}^{\prime}}{a}-{\gamma}^{\prime}{\omega}^{\prime}) = \frac{1}{2}(-e^{-2\lambda}{{\phi}^{\prime}}^2 + 
e^{-2\gamma}{\dot{\phi}}^2)- V(\phi),
\ee
\be
e^{-2\lambda}(2\frac{{a}^{\prime\prime}}{a}+\frac{{{a}^{\prime}}^2}{a^2}+{{\omega}^{\prime}}^2-2{\lambda}^{\prime}\frac{{a}^{\prime}}{a})
-e^{-2\gamma}(2\dot{\lambda}\frac{\dot{a}}{a}+\frac{{\dot{a}}^2}{a^2}-{\dot{\omega}}^2)=\frac{1}{2}(-e^{-2\lambda}{{\phi}^{\prime}}^2 - 
e^{-2\gamma}{\dot{\phi}}^2)- V(\phi),
\ee
\be
-2(\frac{{\dot{a}}^{\prime}}{a}+{\omega}^{\prime}\dot{\omega}-\dot{\lambda}\frac{{a}^{\prime}}{a}-{\gamma}^{\prime}\frac{\dot{a}}{a})=\dot{\phi}{\phi}^{\prime},
\ee
where we have taken $8{\pi}G=1$, a dot and a prime represent differentiations with respect to time `T' and space coordinate `R' respectively. Equations $(2) -(6)$ represent ${G_{1}}^1$, ${G_{2}}^2$, ${G_{3}}^3$, ${G_{0}}^0$ and ${G_{1}}^0$ equations repectively.
\par In order to make the equation system tractable, we assume $\gamma(R,T)$, $\lambda(R,T)$ and $a(R,T)$ are all separable as products of functions 
of $R$ and that of $T$. The $T$-dependence of $\gamma$ and $R$-dependence of $\lambda$ can now be absorbed in $dT^2$ and $dR^2$ respectively by suitable scaling of $T$ and $R$ without any further loss of generality. Now one has, $$\gamma=\gamma(R),$$ $$\lambda=\lambda(T),$$ $$a(R,T)=\alpha(R)\beta(T).$$

\section{Planar symmetry $(\omega=0, f=1)$}
For $\omega=0$ the field equations take the form,
\bea
e^{-2\lambda}(2{\gamma}^{\prime}\frac{{\alpha}^{\prime}}{\alpha}+\frac{{{\alpha}^{\prime}}^2}{{\alpha}^2})-e^{-2\gamma}(2\frac{\ddot{\beta}}{\beta}+
\frac{{\dot{\beta}}^2}{{\beta}^2})=\frac{1}{2}e^{-2\gamma}{\dot{\phi}}^2- V(\phi),\\
e^{-2\lambda}(\frac{{\alpha}^{\prime\prime}}{\alpha}+{\gamma}^{\prime\prime}+{{\gamma}^{\prime}}^2+{\gamma}^{\prime}\frac{{\alpha}^{\prime}}{\alpha})+
e^{-2\gamma}(-\ddot{\lambda}-{\dot{\lambda}}^2-\frac{\ddot{\beta}}{\beta}-\dot{\lambda}\frac{\dot{\beta}}{\beta})=\frac{1}{2}e^{-2\gamma}{\dot{\phi}}^2- V(\phi),\\
e^{-2\lambda}(2\frac{{\alpha}^{\prime\prime}}{\alpha}+\frac{{{\alpha}^{\prime}}^2}{{\alpha}^2})-e^{-2\gamma}(2\dot{\lambda}\frac{\dot{\beta}}{\beta}
+\frac{{\dot{\beta}}^2}{{\beta}^2})=-\frac{1}{2}e^{-2\gamma}{\dot{\phi}}^2- V(\phi),\\
-2\frac{\dot{\beta}}{\beta}\frac{{\alpha}^{\prime}}{\alpha}+2\dot{\lambda}\frac{{\alpha}^{\prime}}{\alpha}+2{\gamma}^{\prime}\frac{\dot{\beta}}{\beta}=0.
\eea
Integrating equation$(10)$ one gets,\\
\be
\frac{\lambda}{\ln{\beta}}=1-\frac{\gamma}{\ln{\alpha}},
\ee
where a constant of integration is put equal to zero.
From equations $(8)$ and $(9)$ we get,
\be
V(\phi)=-\frac{1}{2}e^{-2\lambda}(3\frac{{\alpha}^{\prime\prime}}{\alpha}+\frac{{{\alpha}^{\prime}}^2}{{\alpha}^2}+{\gamma}^{\prime\prime}+{{\gamma}^{\prime}}^2+
{\gamma}^{\prime}\frac{{\alpha}^{\prime}}{\alpha})-\frac{1}{2}e^{-2\gamma}(-\ddot{\lambda}-{\dot{\lambda}}^2-3\dot{\lambda}\frac{\dot{\beta}}{\beta}-\frac{\ddot{\beta}}{\beta}-\frac{{\dot{\beta}}^2}{{\beta}^2}),
\ee
and\\
\be
{\dot{\phi}}^2=e^{-2\lambda}e^{2\gamma}(-\frac{{\alpha}^{\prime\prime}}{\alpha}-\frac{{{\alpha}^{\prime}}^2}{{\alpha}^2}+{\gamma}^{\prime\prime}+{{\gamma}^{\prime}}^2+{\gamma}^{\prime}\frac{{\alpha}^{\prime}}{\alpha})-\ddot{\lambda}-{\dot{\lambda}}^2-\frac{\ddot{\beta}}{\beta}+\dot{\lambda}\frac{\dot{\beta}}{\beta}
+\frac{{\dot{\beta}}^2}{{\beta}^2}.
\ee
In the above system of equations $\phi$ is assumed to be a function of $T$ alone. This is a simplifying assumption made to solve the system of equations easily. Therefore what follows from the above assumption is that $V(\phi)$ is also a function of $T$ and to implement this, we put $\gamma=constant$. The value of the constant is taken as zero without any further loss of generality. Equations $(12)$ and $(13)$ then give,
\bea
3\frac{{\alpha}^{\prime\prime}}{\alpha}+\frac{{{\alpha}^{\prime}}^2}{{\alpha}^2}=k_{1},\\
-\frac{{\alpha}^{\prime\prime}}{\alpha}-\frac{{{\alpha}^{\prime}}^2}{{\alpha}^2}=-k_{2},
\eea
where $k_{1}$ and $k_{2}$ are constants. Both the equations $(14)$ and $(15)$are satisfied by the equation
\be
 \alpha=e^{\frac{\sqrt{k_{1}}}{2}R},
\ee 
and $k_{1}$ and $k_{2}$ are related as $k_{1}=2k_{2}$ for the sake of consistency. The above form of the solution of $\alpha(R)$ also satisfies the 
other equations $(7)$ - $(10)$. With $\gamma=0$, equation $(11)$ therefore has the form
\be
\lambda=\ln{\beta}.
\ee
 With this the metric takes the form $ds^2=dT^2-{\beta}^2d{R}^2-{\beta}^2e^{\sqrt{k_{1}}R}(d{\eta}^2+d{\xi}^2)$ 
and we are still to find $\beta$.
\par We make an assumption that the potential is directly proportional to the kinetic energy term,  i.e.,
\be V(\phi)=m{\dot{\phi}}^2, 
\ee 
where $m$ is a constant. As $g_{00}=1$ and $\phi=\phi(t)$, ${\dot{\phi}}^2$ is actually ${\phi}^{,\mu}{\phi}_{,\mu}$ which is indeed a scalar and equation $(18)$ is thus consistent. However, it deserves mention that the field in this case is not a k-essence field where the Lagrangian is given by $L=L(X)$ where $X={\phi}_{,\mu}{\phi}^{,\mu}$. With the choice of $\gamma=0$, the energy momentum tensor for present model looks like 

$$T^{\mu}_{\nu}= (\dot{\phi})^{2} diag ((1/2 + m), (1/2- m), (1/2- m),(1/2- m)).$$

Actually the model does have a potential, and the present work deals with those which incidentally are proportional to the kinetic part. For instance, the simple case where $\phi$ is spatially homogeneous, a choice like 
$$V(\phi)\propto{exp(\frac{\phi}{\phi _{0}})}$$ will in fact be a possibility that gives rise to equation (18) at least in a special case.

\par The equation of state parameter, $w=\frac{p_{\phi}}{\rho_{\phi}}=\frac{\frac{1}{2}{\dot{\phi}}^2-V(\phi)}{\frac{1}{2}{\dot{\phi}}^2+V(\phi)}$ will be a constant given by $w=\frac{1-2m}{1+2m}$. For any finite positive value of $m$, $w$ remains greater than $-1$ and hence the scalar field is in fact a
 quintessence field.
\par
\begin{figure}[!h]
\centerline{\psfig{figure = 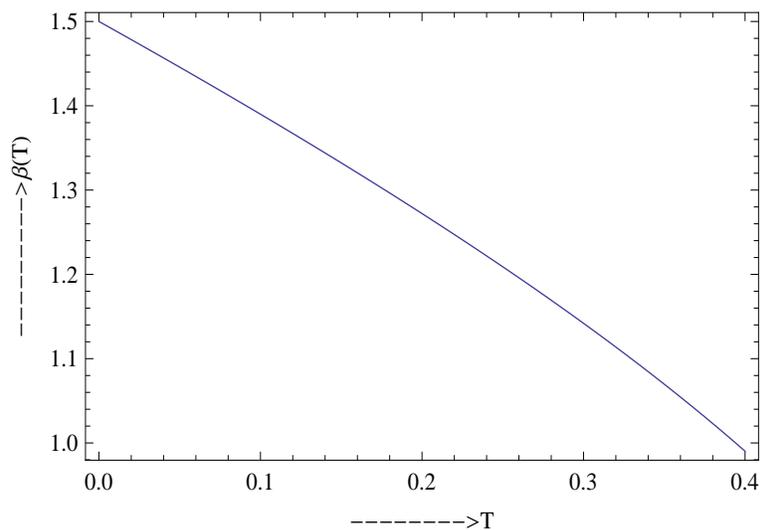, height=70mm, width=100mm}}
\caption{\normalsize{Plot of $\beta(T)$ vs $T$ for $m=0$ with $k_{1}=3$ and $C_{3}=2$ ($\omega=0$)}}
\label{fig1}
\end{figure}

\begin{figure}[!h]
\centerline{\psfig{figure = 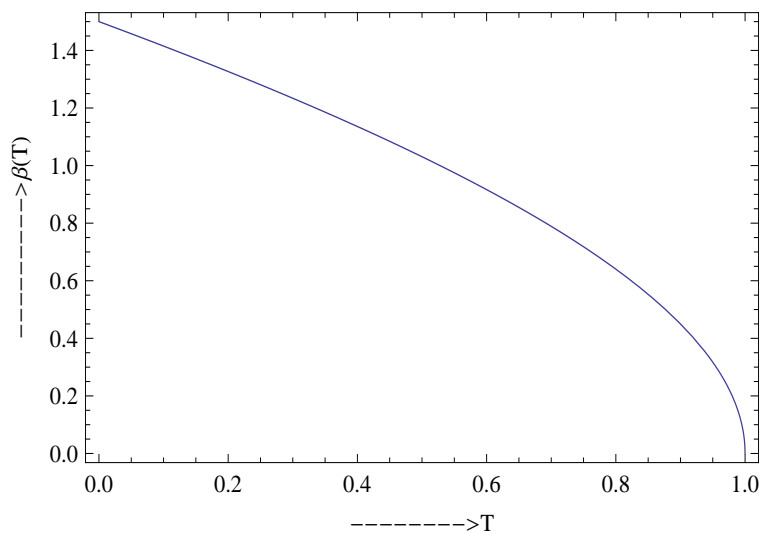, height=70mm, width=100mm}}
\caption{\normalsize{Plot of $\beta(T)$ vs $T$ for $m=1/4$ with $k_{1}=1$ and $C_{3}=1$ ($\omega=0$)}}
\label{fig2}
\end{figure}

 Putting the expressions of $V(\phi)$ and ${\dot{\phi}}^2$ from $(12)$ and $(13)$ one gets,
\be
{\dot{\beta}}^2=\frac{C_{2}}{C_{1}}+C_{3}{\beta}^{-2C_{1}},
\ee
where $C_{1}=\frac{2(1-m)}{1+2m}$, $C_{2}=\frac{k_{1}(1-m)}{2(1+2m)}$ and $C_{3}$ an integration constant. Now since it becomes difficult to further 
integrate equation $(19)$ to obtain $\beta(T)$, we solve it numerically. From $(19)$ we have,
\be
\dot{\beta}=-\sqrt{\frac{C_{2}}{C_{1}}+C_{3}{\beta}^{-2C_{1}}}.
\ee
As we are interested in a collapsing model, the negative root is considered. $\beta(T)$ is plotted as 
a function of $T$ for different values of $m$. For $m=0$ i.e. a scalar field model without any potential, we have 
$\dot{\beta}=-\sqrt{\frac{k_{1}}{4}+C_{3}{\beta}^{-4}}$ and giving some appropriate values of the constants $k_{1}$ and $C_{3}$ and applying reasonable
 boundary conditions it is seen from (figure \ref{fig1}) that $\beta(T)$ starts from a finite value but becomes zero at a finite time. As the proper volume is proportional to ${\beta}^3$ $(\sqrt{-g}={\alpha}^2{\beta}^3$), this indicates that the distribution collapses to the singularity of a zero proper volume. Different reasonable sets of values for $k_{1}$ and $C_{3}$ would lead to different values for time $T$ when the singularity is attained but the nature of the collapse remains similar.\\
  For $m=1/4$ equation $(19)$ leads to $\dot{\beta}=-\sqrt{\frac{k_{1}}{4}+C_{3}{\beta}^{-2}}$. Here also we have a collapsing
model for different values of the constant (figure \ref{fig2} shows one of them).
\par
\begin{figure}[!h]
\centerline{\psfig{figure = 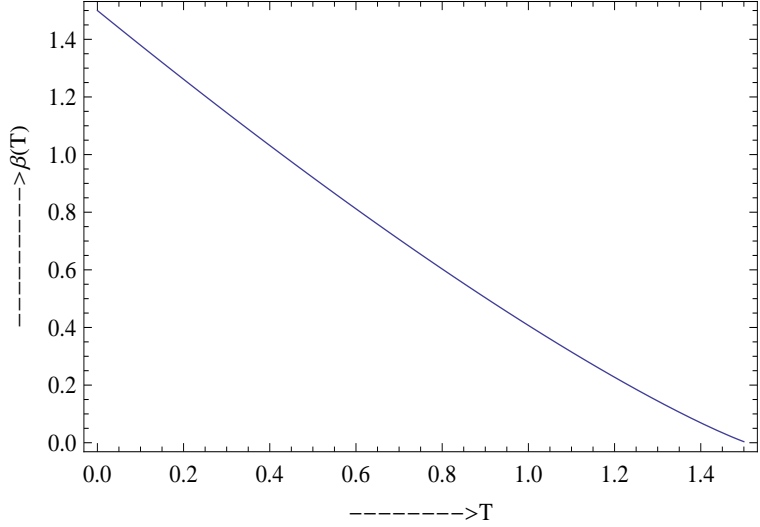, height=70mm, width=100mm}}
\caption{\normalsize{Plot of $\beta(T)$ vs $T$ for $m=3/2$ with $k_{1}=1$ and $C_{3}=1$ ($\omega=0$)}}
\label{fig3}
\end{figure}

 For these cases, a collapsing model is not surprising as the matter content satisfies the energy conditions $(\rho +3p>0)$. If $m>1/2$, the effective pressure becomes negative and when $m>1$, the energy condition is violated and the scalar field is apt to act as a dark energy. As an example we choose $m=3/2$, for which equation $(19)$ yields $\dot{\beta}=-\sqrt{\frac{k_{1}}{4}+C_{3}{\beta}^{1/2}}$. With some reasonable  boundary conditions and putting the values of constants $k_{1}$ and $C_{3}$ both equal to $1$ one can get a collapsing model as seen in (figure \ref{fig3}).
\par For values of $m$ less than $1/2$ we have sharply collapsing models where the volume falls to zero very rapidly particularly towards the end. For $m=3/2$, the rate of collapse slightly flattens out towards the end, but one can have a collapsing model none-the-less.
\par It is relevant to point out that at the singularity $(\beta(T)\rightarrow 0)$, both ${\dot{\phi}}^2$ and $V(\phi)$ would blow up leading to an infinite energy density and pressure. Equation (13) now yields
$$\dot{\phi}=({\phi}_{1}{\beta}^{-2} +{\phi}_{2}{\beta}^{-n})^{\frac{1}{2}}$$ where ${\phi}_{1}$, ${\phi}_{2}$ are positive constants, related with the constants of integration and $m$, and 
$n=6/(1+2m).$ This cannot be analytically integrated for all values of $m$. If one chooses $m=1$, it can be shown that $\phi \propto \ln{T}$. Using the ansatz (18), one can now easily show that 

                             $$V(\phi)=me^{-2\phi}.$$

\section{Cylindrical, toroidal, pseudoplanar symmetry ($\omega\ne 0$)}
 Now some models with a non-zero $\omega$ is considered which would lead to cylindrical, toroidal or pseudoplanar kind of symmetry depending on the range of the coordinates. When the coordinates 
$\eta$ and $\xi$range from $-\infty$ to $+\infty$, one has a pseudoplanar symmetry. If one of the coordinates $\eta$ or $\xi$ is an angle coordinate ranging from $0$ to $2\pi$, one would have a cylindrical symmetry. If both of them are angles, one has a toroidal symmetry. In what follows, we assume $\omega$ is either a function of $R$ or that of $T$ for the sake of simplicity.
\subsection{$\omega=\omega(R)$}
With $\omega=\omega(R)$ and the same simplifications as before, that is, $\gamma=\gamma(R)$, $\lambda=\lambda(T)$, $a(R,T)=\alpha(R)\beta(T)$ and $\phi=\phi(T)$ the field equations are as follows,
\be
e^{-2\lambda}(2{\gamma}^{\prime}\frac{{\alpha}^{\prime}}{\alpha}+\frac{{{\alpha}^{\prime}}^2}{{\alpha}^2}-{{\omega}^{\prime}}^2)-e^{-2\gamma}(2\frac{\ddot{\beta}}{\beta}+
\frac{{\dot{\beta}}^2}{{\beta}^2})=\frac{1}{2}e^{-2\gamma}{\dot{\phi}}^2- V(\phi),
\ee
\be
e^{-2\lambda}(\frac{{\alpha}^{\prime\prime}}{\alpha}+2{\omega}^{\prime}\frac{{\alpha}^{\prime}}{\alpha}+{{\omega}^{\prime}}^2+{\omega}^{\prime\prime}+{\gamma}^{\prime\prime}+{{\gamma}^{\prime}}^2+{\gamma}^{\prime}\frac{{\alpha}^{\prime}}{\alpha}+{\gamma}^{\prime}{\omega}^{\prime})+
e^{-2\gamma}(-\ddot{\lambda}-{\dot{\lambda}}^2-\frac{\ddot{\beta}}{\beta}-\dot{\lambda}\frac{\dot{\beta}}{\beta})=\frac{1}{2}e^{-2\gamma}{\dot{\phi}}^2-
 V(\phi),
\ee
\be
e^{-2\lambda}(\frac{{\alpha}^{\prime\prime}}{\alpha}-2{\omega}^{\prime}\frac{{\alpha}^{\prime}}{\alpha}+{{\omega}^{\prime}}^2-{\omega}^{\prime\prime}+{\gamma}^{\prime\prime}+{{\gamma}^{\prime}}^2+{\gamma}^{\prime}\frac{{\alpha}^{\prime}}{\alpha}-{\gamma}^{\prime}{\omega}^{\prime})+
e^{-2\gamma}(-\ddot{\lambda}-{\dot{\lambda}}^2-\frac{\ddot{\beta}}{\beta}-\dot{\lambda}\frac{\dot{\beta}}{\beta})=\frac{1}{2}e^{-2\gamma}{\dot{\phi}}^2-
 V(\phi),
\ee
\bea
e^{-2\lambda}(2\frac{{\alpha}^{\prime\prime}}{\alpha}+\frac{{{\alpha}^{\prime}}^2}{{\alpha}^2}+{{\omega}^{\prime}}^2)-e^{-2\gamma}(2\dot{\lambda}\frac{\dot{\beta}}{\beta}
+\frac{{\dot{\beta}}^2}{{\beta}^2})=-\frac{1}{2}e^{-2\gamma}{\dot{\phi}}^2- V(\phi),\\
-2\frac{\dot{\beta}}{\beta}\frac{{\alpha}^{\prime}}{\alpha}-2{\omega}^{\prime}\dot{\omega}+2\dot{\lambda}\frac{{\alpha}^{\prime}}{\alpha}+2{\gamma}^
{\prime}\frac{\dot{\beta}}{\beta}=0.
\eea
Since $\omega=\omega(R)$ only, equation $(25)$ takes the form,
\be
-2\frac{\dot{\beta}}{\beta}\frac{{\alpha}^{\prime}}{\alpha}+2\dot{\lambda}\frac{{\alpha}^{\prime}}{\alpha}+2{\gamma}^{\prime}\frac{\dot{\beta}}{\beta}=0.
\ee
Integrating this equation we get
\be
\frac{\lambda}{\ln{\beta}}=1-\frac{\gamma}{\ln{\alpha}},
\ee
exactly the same relation as in equation $(11)$. By the same logic of $\phi$ being a function of $T$ only, $\gamma$ is a constant and is chosen to be 
zero. Equations $(22)$ and $(23)$ yield an equation of the form
\be
\frac{{\omega}^{\prime\prime}}{{\omega}^{\prime}}+2\frac{{\alpha}^{\prime}}{\alpha}=0,
\ee
which on integrating gives,
\be
{\omega}^{\prime}=\frac{A}{{\alpha}^2},
\ee 
 where $A$ is a constant of integration. Substituiting ${\omega}^{\prime}$ from $(29)$ in the field equations, one has an equation of the form,
\be
\frac{{{\alpha}^{\prime}}^2}{{\alpha}^2}-\frac{A^2}{{\alpha}^4}=K,
\ee 
where $K$ is a constant. Integrating the above equation one gets the solution of $\alpha(R)$ as 
\be
\alpha(R)=A_{1}\sqrt{\sinh(R+C)}
\ee
where $A_{1}=\sqrt{\frac{A}{K}}$ and $C$ is an integration constant and the solution is consistent with all the field equations. 
Now putting the solution of $\alpha(R)$ from $(31)$ and integrating equation $(29)$ one obtains 
\be
\omega=K\ln\left(\tanh\frac{R+C}{2}\right).
\ee
Using equations $(21)$ and $(24)$ we have
\bea
V(\phi)=2\frac{{\dot{\beta}}^2}{{\beta}^2}+\frac{\ddot{\beta}}{\beta}-\frac{K}{{\beta}^2},\\
{\dot{\phi}}^2=2\frac{{\dot{\beta}}^2}{{\beta}^2}-2\frac{\ddot{\beta}}{\beta}-2\frac{K}{{\beta}^2}.
\eea
 Using equations $(33)$, $(34)$ and the ansatz $(18)$ one can write,
\be
{\dot{\beta}}^2=\frac{K(1-2m)}{2(1-m)}+E\beta^{\frac{-4(1-m)}{1+2m}}
\ee
where $E$ is an integration constant. Taking the positive root of $\dot{\beta}$ gives an ever expanding model. But considering the negative root
there is a possibility to have a collapsing model and hence the negative root is chosen. 
\par
\begin{figure}[!h]
\centerline{\psfig{figure = 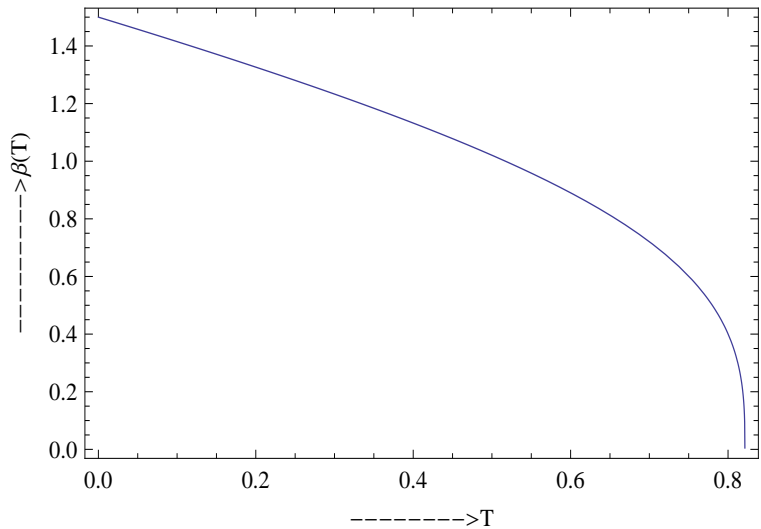, height=70mm, width=100mm}}
\caption{\normalsize{Plot of $\beta(T)$ vs $T$ for $m=0$ with $K=1$ and $E=1$ ($\omega=\omega(R)$)}}
\label{fig4}
\end{figure}

\begin{figure}[!h]
\centerline{\psfig{figure = 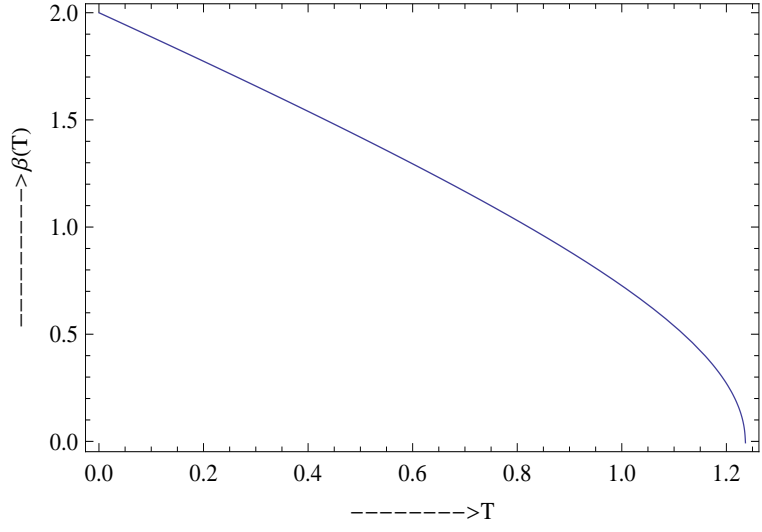, height=70mm, width=100mm}}
\caption{\normalsize{Plot of $\beta(T)$ vs $T$ for $m=1/4$ with $K=1$ and $E=1$ ($\omega=\omega(R)$)}}
\label{fig5}
\end{figure}

\begin{figure}[!h]
\centerline{\psfig{figure = 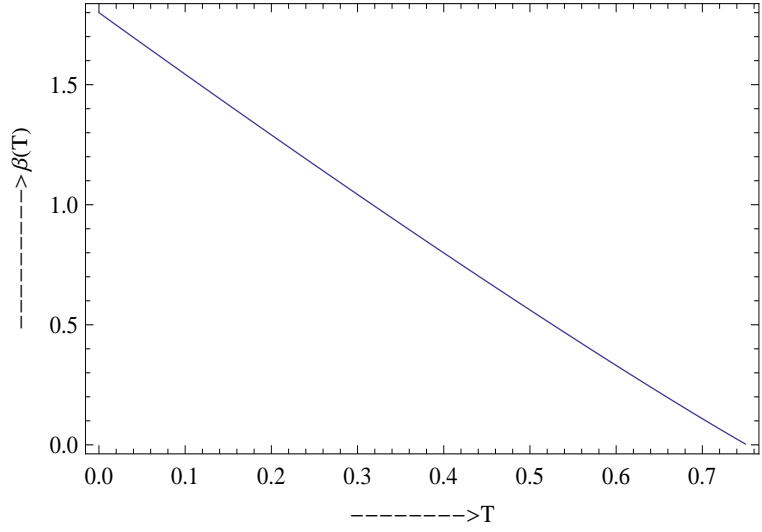, height=70mm, width=100mm}}
\caption{\normalsize{Plot of $\beta(T)$ vs $T$ for $m=3/2$ with $K=4$ and $E=2$ ($\omega=\omega(R)$)}}
\label{fig6}
\end{figure}

 Putting the value of $m=0$ and constants $E$ and $K$ equal to $1$, equation $(35)$ looks like $\dot{\beta}=-\sqrt{\frac{K}{2}+E{\beta}^{-4}}$. From the numerical plot of $\beta(T)$ vs $T$ in (figure \ref{fig4}) it is seen that the volume starts from non-zero value at some initial time $T=0$ and then becomes zero at a finite value of time $T$. Putting some other values of the constants only shifts the time at which the proper volume crushes to the singularity but the nature of the graphs remain unchanged.\\
A value of $m=1/4$ that ie $V(\phi)=\frac{1}{4}{\dot{\phi}^2}$ and same values of the constants also yield a model collapsing in nature as is evident from (figure \ref{fig5}).\\
 As before we take same example of $m=3/2$ to represent a dark energy candidate. Equation $(35)$ now becomes $\dot{\beta}=-\sqrt{2K+E{\beta}^{1/2}}$. It is evident from the plot in (figure \ref{fig6}) that the volume becomes zero gradually at some value of $T$. In this case also the different values of the constants only change the time for which $\beta(T)\rightarrow 0$ keeping the nature of the graphs unchanged. \\

Equation (34) and (35) yield 

$$\dot{\phi}=({\phi}_{3}{\beta}^{-2} +{\phi}_{4}{\beta}^{-l})^{\frac{1}{2}}$$ where ${\phi}_{3}$, ${\phi}_{4}$ are constants, related with the constants of integration and $m$, and $l=6/(1+2m)$. This cannot in general be integrated. But $\dot{\phi}$ indeed diverges when $\beta$, and hence the proper volume, goes to zero. Thus the energy denisty has a singularity when $\beta$ goes to zero.

\subsection{$\omega=\omega(T)$}
\begin{figure}[!h]
\centerline{\psfig{figure = 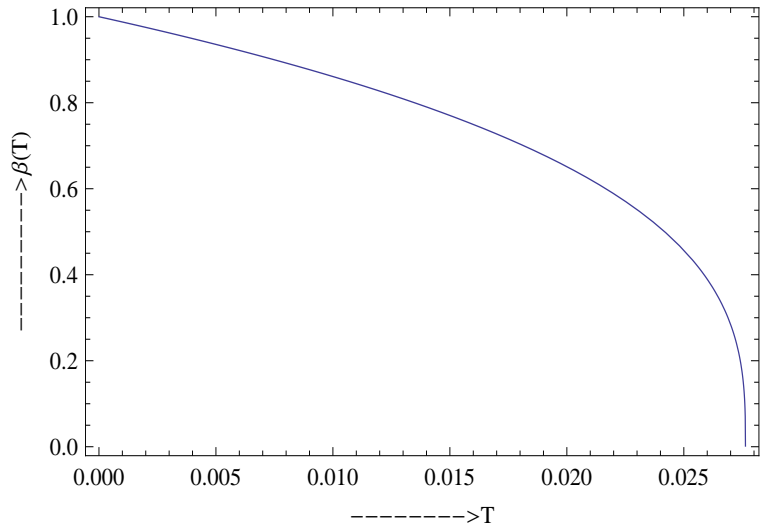, height=70mm, width=100mm}}
\caption{\normalsize{Plot of $\beta(T)$ vs $T$ for $m=0$ with $K_{1}=0.002$, $B=9$ and $D=150$ ($\omega=\omega(T)$)}}
\label{fig7}
\end{figure}

\begin{figure}[!h]
\centerline{\psfig{figure = 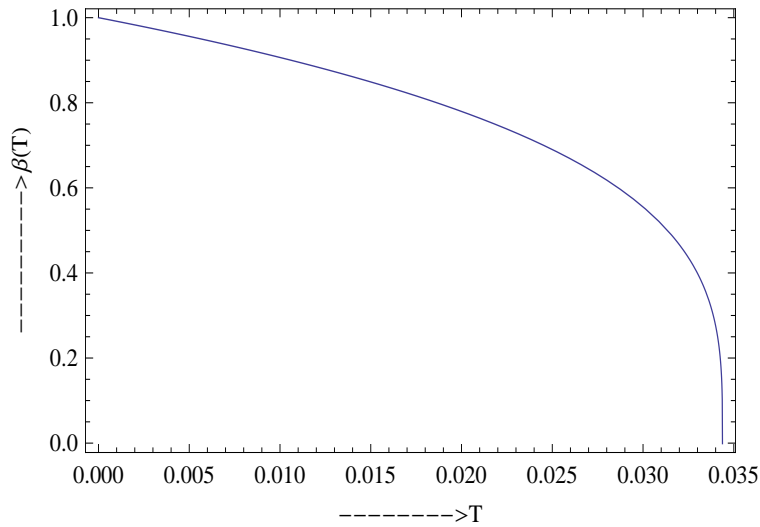, height=70mm, width=100mm}}
\caption{\normalsize{Plot of $\beta(T)$ vs $T$ for $m=1/4$ with $K_{1}=6$, $B=9$ and $D=10$ ($\omega=\omega(T)$)}}
\label{fig8}
\end{figure}

\begin{figure}[!h]
\centerline{\psfig{figure = 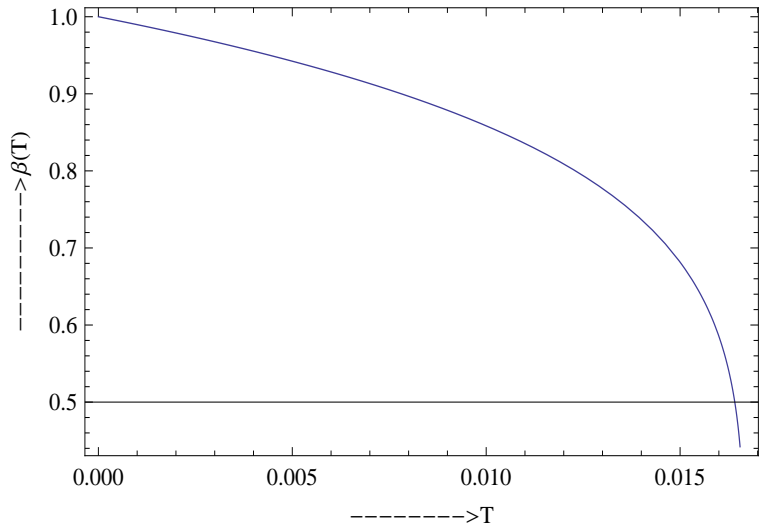, height=70mm, width=100mm}}
\caption{\normalsize{Plot of $\beta(T)$ vs $T$ for $m=3/2$ with $K_{1}=-0.002$, $B=0.2$ and $D=100$ ($\omega=\omega(T)$)}}
\label{fig9}
\end{figure}

Taking $\omega=\omega(T)$ the field equations can be written as,
\be
e^{-2\lambda}(2{\gamma}^{\prime}\frac{{\alpha}^{\prime}}{\alpha}+\frac{{{\alpha}^{\prime}}^2}{{\alpha}^2})-e^{-2\gamma}(2\frac{\ddot{\beta}}{\beta}+
\frac{{\dot{\beta}}^2}{{\beta}^2}+{\dot{\omega}}^2)=\frac{1}{2}e^{-2\gamma}{\dot{\phi}}^2- V(\phi),
\ee
\be
e^{-2\lambda}(\frac{{\alpha}^{\prime\prime}}{\alpha}+{\gamma}^{\prime\prime}+{{\gamma}^{\prime}}^2+{\gamma}^{\prime}\frac{{\alpha}^{\prime}}{\alpha})+
e^{-2\gamma}(-\ddot{\lambda}-{\dot{\lambda}}^2-\frac{\ddot{\beta}}{\beta}-2\frac{\dot{\beta}}{\beta}\dot{\omega}-{\dot{\omega}}^2-\ddot{\omega}-\dot{\lambda}\frac{\dot{\beta}}{\beta}-\dot{\lambda}\dot{\omega})=\frac{1}{2}e^{-2\gamma}{\dot{\phi}}^2- V(\phi),
\ee
\be
e^{-2\lambda}(\frac{{\alpha}^{\prime\prime}}{\alpha}+{\gamma}^{\prime\prime}+{{\gamma}^{\prime}}^2+{\gamma}^{\prime}\frac{{\alpha}^{\prime}}{\alpha})
e^{-2\gamma}(-\ddot{\lambda}-{\dot{\lambda}}^2-\frac{\ddot{\beta}}{\beta}+2\frac{\dot{\beta}}{\beta}\dot{\omega}-{\dot{\omega}}^2+\ddot{\omega}-\dot{\lambda}\frac{\dot{\beta}}{\beta}+\dot{\lambda}\dot{\omega})=\frac{1}{2}e^{-2\gamma}{\dot{\phi}}^2- V(\phi),
\ee
\be
e^{-2\lambda}(2\frac{{\alpha}^{\prime\prime}}{\alpha}+\frac{{{\alpha}^{\prime}}^2}{{\alpha}^2})-e^{-2\gamma}(2\dot{\lambda}\frac{\dot{\beta}}{\beta}
+\frac{{\dot{\beta}}^2}{{\beta}^2}-{\dot{\omega}}^2)=-\frac{1}{2}e^{-2\gamma}{\dot{\phi}}^2- V(\phi),
\ee
\be
-2\frac{\dot{\beta}}{\beta}\frac{{\alpha}^{\prime}}{\alpha}+2\dot{\lambda}\frac{{\alpha}^{\prime}}{\alpha}+2{\gamma}^
{\prime}\frac{\dot{\beta}}{\beta}=0,
\ee
 the same simplifying assumptions like $\gamma=\gamma(R)$, $\lambda=\lambda(T)$, $a(R,T)=\alpha(R)\beta(T)$ and $\phi=\phi(T)$ are kept intact. Following similar arguments as before, one can write, $\lambda=\ln\beta$. Solving the above field equations $(36)$ to $(39)$ one easily arrives at
\be
\alpha={e}^{\sqrt{K_{1}}R},
\ee
which is consistent with all the field equations, $K_{1}$ being a constant of integration. Now subtracting equation $(38)$ from $(37)$ one obtains after integration,
\be
\dot{\omega}=\frac{B}{{\beta}^3},
\ee
where $B$ is a constant of integration.
Equations $(36)$ and $(39)$ give,
\be
V(\phi)=(\frac{\ddot{\beta}}{\beta}+2\frac{{\dot{\beta}}^2}{{\beta}^2})-e^{-2\lambda}(\frac{{\alpha}^{\prime\prime}}{\alpha}+\frac{{{\alpha}^{\prime}}^2}{{\alpha}^2}),
\ee
and
\be
{\dot{\phi}}^2=-2(\frac{\ddot{\beta}}{\beta}+{\dot{\omega}}^2+\frac{{\dot{\beta}}^2}{{\beta}^2})-2e^{-2\lambda}\frac{{\alpha}^{\prime\prime}}{\alpha}.
\ee
Use of equations $(17), (41)$ and $(42)$ in equations $(43)$ and $(44)$ results in 
\be
V(\phi)= \frac{\ddot{\beta}}{\beta}+2\frac{{\dot{\beta}}^2}{{\beta}^2}-2\frac{K_{1}}{{\beta}^2},
\ee
and
\be
{\dot{\phi}}^2=-2(\frac{\ddot{\beta}}{\beta}+\frac{{B}^2}{{\beta}^6}+\frac{{\dot{\beta}}^2}{{\beta}^2})-2\frac{K_{1}}{{\beta}^2}.
\ee
Making the same assumption as in equation $(18)$ and integrating a combination of equations $(45)$ and $(46)$ one gets,
\be
{\dot{\beta}}^2=\frac{K_{1}(1-m)}{2(1+m)}-\frac{B}{2}{\beta}^{-4}+D{\beta}^{-4(1+m)},
\ee
where $D$ is a constant of integration.
Choosing the negative root of $\beta(T)$, the plots of $\beta(T)$ vs $T$ reveal that for $m=0$ and $m=1/4$, with appropriate values of the constants, the volume crushes to zero sharply at some finite time (figure \ref{fig7} and figure \ref{fig8}).
Plotting $\beta(T)$ vs $T$ for $m=3/2$ also, with suitable value of the constants, one gets a collapsing model (figure \ref{fig9}).
 The point to be noted is that in the previous models for $\omega=0$ or $\omega=\omega(R)$, $m=3/2$ i.e., when the scalar field is consistent with a dark energy candidate, we had collapsing models where the volume becomes zero in a sedate manner, but in this case when $\omega=\omega(T)$ even for values of $m>1/2$ we have collapsing models where the volume falls to zero quite abruptly.
\par In this case also one can write $\dot{\phi}$ as a pwer function of $\beta$ as 

$$\dot{\phi}=({\phi}_{5}{\beta}^{-2} +{\phi}_{6}{\beta}^{-6} +{\phi}_{7}{\beta}^{-4(1+m)})^{\frac{1}{2}},$$

where ${\phi}_{5}$, ${\phi}_{6}$ and ${\phi}_{7}$ are constants, related with the constants of integration and $m$. One cannot intergrate for ${\phi}$ except for very special cases. Like the other cases, one still has a divergence of energy density at the singularity of zero proper volume.

\section{Discussions}
Non spherically symmetric matter distributions with a minimally coupled scalar field can lead to collapsing models. With some choices of the arbitrary constants of integration, the models evolve to produce singularity of a zero proper volume at some finite future. The intriguing feature is that even for the cases where the scalar field acts as a dark energy candidate, one can have collapsing solutions with the only difference that for these cases the volume gradually becomes zero unlike the cases where $m<1/2$ which leads to a suddenly collapsing situation at a finite future. However, in one case, $\omega=\omega(T)$, even the dark energy candidates may lead to a sudden collapse. For a positive value of $m$, it is obvious that $w>-1$, that is, the model will never be a phantom. It infact mimics a quintessence field.

\par This work is indeed a specal case where the potential is assumed to be proportional to the kinetic part, but actually the model does not mimic the k-essence model. Here one does have a potential. At least in one special case, for the planar symmetry, an example of the potential is given which can give rise to this kind of a behaviour.

\par This work also clearly indicates that the clustering properties of a `dark energy' indeed deserves a more critical attention.
\vskip .5in
%\pagebreak

{\large\bf Acknowledgement:}\\

\par The authors thank the the anonymous referee for suggesting some important improvement and the BRNS (DAE) for financial support.\\

\end{document}